\documentclass[universe,article,accept,pdftex,oneauthor]{Definitions/mdpi} 

\firstpage{1} 
\pubvolume{1}
\issuenum{1}
\articlenumber{0}
\pubyear{2023}
\copyrightyear{2023}
\externaleditor{Academic Editor: E. DiValentino, L. Perivolaropoulos and J. Said}

\datereceived{31 May 2023} 
\daterevised{6 July 2023}
\dateaccepted{20 July 2023} 
\datepublished{ } 
\hreflink{https://doi.org/} 

\Title{Untying the Growth Index to Relieve the $\sigma_8$ Discomfort}

\TitleCitation{Untying the Growth Index to Relieve the $\sigma_8$ Discomfort}

\Author{Ziad Sakr $^{1,2,3}$\orcidA{}}

\AuthorNames{Ziad Sakr}

\AuthorCitation{Sakr, Z.}
\address{%
$^{1}$ \quad Institut f\"ur Theoretische Physik, University of Heidelberg, Philosophenweg 16, 69120 Heidelberg, Germany; ziad.sakr@net.usj.edu.lb\\
$^{2}$ \quad IRAP, Universit\'e de Toulouse, CNRS, CNES, UPS, 31400 Toulouse, France\\
$^{3}$ \quad Faculty of Sciences, Universit\'e St Joseph, 17-5208 Beirut,Lebanon}

\abstract{The matter fluctuation parameter $\sigma_8$ is, by model construction, degenerate with the growth index $\gamma$. Here, we study the effect on the cosmological parameter constraints by treating each independently from one another, considering $\sigma_8$ as a free and non-derived parameter along with a free $\gamma$. We then try to constrain all parameters using three probes that span from deep to local redshifts, namely the CMB spectrum, the growth measurements from redshift space distortions $f\sigma_8$, and the galaxy cluster counts. We also aim to assess the impact of this relaxation on the $\sigma_8$ tension between its inferred CMB value in comparison to that obtained from local cluster counts. We also propose a more sophisticated correction, along with the classical one, that takes into account the impact of cosmology on the growth measurements when the parameters are varied in the Monte Carlo process, which consist in adjusting the growth to keep the observed power spectrum, integrated over all angles and scales, as invariant with the background evolution. We found by using the classical correction that untying the two parameters does not shift the maximum likelihood of either $\sigma_8$ or $\gamma$, but it rather enables larger bounds with respect to when $\sigma_8$ is a derived parameter, and that when considering CMB + $f\sigma_8$, or when further combining with cluster counts albeit with tighter bounds. Precisely, we obtain $\sigma_8 = 0.809\pm 0.043 $ and $\gamma = 0.613\pm 0.046$ in agreement with Planck's constraint for the former and compatible with $\Lambda$CDM for the latter but with bounds wide enough to accommodate both values subject to the tensions. Allowing for massive neutrinos does not change the situation much. On the other hand, considering a tiered correction yields $\sigma_8 = 0.734\pm 0.013$ close to $\sim$1~$\sigma$ for the inferred local values albeit with a growth index of $\gamma = 0.636\pm 0.022$ at $\sim$2~$\sigma$ from its $\Lambda$CDM value. Allowing for massive neutrinos in this case yielded $\sigma_8 = 0.756\pm 0.024$, still preferring low values but with much looser constraints on $\gamma = 0.549\pm 0.048$ and a slight preference for $\Sigma m_\nu \sim 0.19$. We conclude that untying $\sigma_8$ and $\gamma$ helps in relieving the discomfort on the former between the CMB and local probes, and that careful analysis should be followed when using data products treated in a model-dependent way.}


\begin{document}

\section{Introduction}

The study of the formation and distribution of large-scale structures (LSS) is one of the strongest tools used to constrain a cosmological model \citep{1992opc..book..171J}, with the essential ingredients necessary to describe it being the matter density, the~matter fluctuation calibration parameter, $\sigma_8$, and the matter growth rate. The~latter can be parameterised as a function of the matter density to the power of what we call the growth index $\gamma$, for which a value of $\sim$0.55 was found to effectively describe the growth in the $\Lambda$CDM model \citep{1980lssu.book.....P}. Deviation from it could signal the need for models beyond the standard cosmological one, e.g.,~\cite{Wen:2023bcj}. 

Among the probes that serve to constrain these parameters, are the ones used in this work, such as the cosmic microwave background (CMB) temperature and polarisation angular power spectrum, since it could be related to the growth of the density fluctuations at the recombination epoch \citep{Hu:1995kot}, or the measurements of growth in the clustering of galaxies obtained from its effect on the redshift space distortions (RSD) \citep{Hamilton:1995py}, or the galaxy cluster counts (CC) from measuring their abundance in a given volume \citep{2012ARA&A..50..353K}.

However, among the three parameters, the CMB only directly measures the matter density, and needs the assumption of a model to derive $\sigma_8$, since what is actually measured is rather the amplitude of the power spectrum $A_s$, extrapolated to our time to obtain the $\sigma_8$ in a model-dependent way. Moreover, the matter density at the recombination epoch is close to unity, resulting in little constraining power on the growth index value. While, for the other two probes, the growth measurements from the galaxy redshift distortions (RSD) and the cluster counts (CC), they actually measure a combination of the three parameters but are not equally sensitive to each of them since the RSD is rather affected by the growth rate of change while the CC is obtained by calibrating a Gaussian halo distribution which average and standard deviation are function of the growth, the~$\sigma_8$ parameter and the matter density at the observed epoch. This results from the fact that, in general, the three parameters are not independent and could be derived from each others if we consider a specific model, such as $\Lambda$CDM \citep{1993MNRAS.262.1023W}.
Moreover, a model is needed for the RSD growth measurements to describe the galaxy bias that relates the underlying matter to the galaxy distribution, while for the cluster counts we need to model the mass-observable scaling relation that relates the mass of the halo to an observed property of the cluster, such as its luminosity, richness, or temperature. 

We can then try to determine the mass-observable calibration parameter for the CCs from hydrodynamical simulations while for the RSD probe, we often marginalise over the bias if we want to obtain the growth alone. Then the degeneracy and dependency between $\Omega_m$, $\sigma_8$, and $\gamma$ could also be alleviated if we do not want to assume an underlying cosmological model, only if we combine all the three probes as we shall see later. However, the need to relax the underlying model has become more relevant lately following findings of a persistent tension, varying from two to four $\sigma$ depending on the datasets used, on~the value of $\sigma_8$ when measured by deep probes, such as the CMB, in comparison to that determined from local probes such as the weak lensing measurements \cite{DES:2021wwk,KiDS:2020ghu} or one of our chosen probe, the CCs \citep{2014A&A...571A..29P}, but also \cite{Benisty:2020kdt,Nunes:2021ipq,Nguyen:2023fip} found a tension on $\sigma_8$ as well between RSD and CMB data. Apart from the possibility that it could result from a mis-determination of the systematics involved, it could also suggest the need for models beyond LCDM in order to cure this ``discomfort''.

Consequently, the growth index $\gamma$ was investigated in~\cite{Sakr:2018new,Ilic:2019pwq} as a way to alleviate this tension, by means of Bayesian studies using a combination of two of the above probes, CMB and CCs, in an agnostic approach, where the parameters in relation to the $\sigma_8$ tension and the calibration parameters were left free to vary. In doing so, \cite{Ilic:2019pwq} showed that even if we let the mass-observable free, we are still able to constrain $\gamma$, preventing it from fully solving the tension even if it reduces it from four to two $\sigma$. However, it was also found that when adding neutrinos to the free $\gamma$ or further relaxing the $\sigma_8$ value, by considering it as a free parameter and not derived from $A_s$, the tension is alleviated albeit with a widening of the constraints, and a fixed $\gamma$ to the $\Lambda$CDM value. Staying with this $\gamma$ restriction, \cite{Blanchard:2021dwr} combined the CMB and CCs with the growth from the RSD measurements and showed that the combination of these three probes prevents a free $\sigma_8$ from solving the tension again; however, as mentioned, they did not vary $\gamma$ but fixed it to its $\Lambda$CDM value. Moreover, they used the growth data obtained assuming $\Lambda$CDM while, in general, when performing an MCMC, a correction accounting for the effects of the~\cite{Alcock:1979mp} (AP) effect, describing the impact of the change of geometry on the growth of structures with the changing cosmology in the Monte Carlo exploration, is applied.

Here we follow the same approach as~\cite{Blanchard:2021dwr} except we first use a larger set of $f\sigma_8$ measurements that span over a large range of redshifts. Second, we also let the growth index vary (see~\cite{{Kazantzidis:2018rnb}} for an RSD CMB assessment of the tension with modified gravity) in addition to $\Omega_m$, $\sigma_8$ and the mass-observable calibration parameter, and end by further letting free the neutrino mass. Finally, we apply the usual AP correction performed when using growth from RSD measured in a specific cosmology. This correction, despite being widely used, assumes that the adjustment is enough independent of the direction of observation and of its scale dependence. This is not completely true, since the growth measurements are usually extracted from an observed power spectrum that includes such effects on all the scales and observed directions. Already \cite{2016MNRAS.456.3743A} noted that a more sophisticated correction needs to be performed, and for that they used a more elaborate method that tried to take into account the direction and the scale over which the power spectrum was measured. However they also assumed small deviations from the fiducial model, sufficient for their purpose since they did not include probes other than the RSD. Here, in addition to the simple AP correction, we shall also test the impact of a more sophisticated one, relaxing the isotropy assumption along with supposing large deviations from the fiducial model, and adopting the ansatz that the observed power spectrum, integrated over all angles and scales, is an invariant quantity when the cosmology changes. Therefore, it can be used to determine the new value of the growth rate by adjusting the latter to keep the integrated observed power spectrum unchanged from the AP effects.

The paper is organised as follows: in Section~\ref{sec:datamodel} we present the pipeline and data used in our analysis, as~well as the model independent approach followed when combining the different datasets, while we show and discuss our results in Section~\ref{sect:results} and conclude in Section~\ref{sect:conclusion}. 

\section{Datasets Treatment and Analysis~Methods}\label{sec:datamodel}

We perform an MCMC Bayesian study using the CMB $C_{\ell}$ of the temperature, polarisation and their cross correlations, from~the publicly available datasets of the Planck mission~\cite{Aghanim:2018eyx} and its likelihood~\cite{Planck:2019nip}.

We combine them with the SZ detected clusters sample PSZ2 containing a total of 439 clusters \citep{2016A&A...594A..27P} spanning the redshift range from $z\sim 0.0$ to $z\sim1.3$, where the distribution of clusters function of redshift and signal-to-noise is written as
\begin{equation}
\frac{{\rm d}N}{{\rm d}z {\rm d}q} = \int {\rm d}\Omega_{\rm mask} \int {\rm d}{M_{500}} \, \frac{{\rm d}N}{{\rm d}z {\rm d}{M_{500}} {\rm d}\Omega}\, P[q | {\bar{q}_{\rm m}}({M_{500}},z,l,b)],
\end{equation}
with
\begin{equation}
\label{eq:dndzdq}
\frac{{\rm d}N}{{\rm d}z {\rm d}{M_{500}} {\rm d}\Omega} = \frac{{\rm d}N}{{\rm d}V {\rm d}{M_{500}}}\frac{{\rm d}V}{{\rm d}z{\rm d}\Omega},
\end{equation}
where the halo mass-function (HMF) can be written in a simple form \citep{1992A&A...264..365B}
\begin{equation}\label{eq:nm0}
  {\rm d} N/{\rm d} m=-\frac{\bar{\rho}}{m}\frac{{\rm d \, ln}\,\nu}{{\rm d \, ln} \,m} \mathcal{F}(\nu) 
\end{equation}
with 
 $\nu=\delta_c/\sigma(M,z)$ where $\sigma(M,z)$, the variance of the linearly evolved density field smoothed by a spherical top-hat window function $W$ of comoving radius $R$ enclosing
mass $M=4\pi\rho_{\rm m}R^3/3$, is:
\begin{equation}
  \sigma^2(M,z) = \frac{1}{2\pi^2} \int_0^\infty k^2 P_m(k,z) |W_{\rm M}(k)|^2 dk\,,
\label{eq:sigR}
\end{equation}
and $\mathcal{F}(\nu)$ the multiplicity function taken from \cite{Despali:2015yla}.\\
The quantity $P[q | {\bar{q}_{\rm m}}({M_{500}},z,l,b)]$ is the distribution of $q$ given the mean signal-to-noise value, ${\bar{q}_{\rm m}}({M_{500}},z,l,b)$, predicted by the model for a cluster of mass ${M_{500}}$ which we relate to the measured integrated Compton $y$-profile $\bar{Y}_{500}$ using the following \mbox{scaling~relations:}
\begin{equation}
\label{eq:Yscaling} 
E^{-\beta}(z)\left[\frac{{D_{\rm A}}^2(z) {\bar{Y}_{500}}}{\mathrm{10^{-4}\,Mpc^2}}\right] =  Y_\ast \left[ {\frac{h}{0.7}}
  \right]^{-2+\alpha} \left[\frac{(1-b)\,
    {M_{500}}}{6\times10^{14}\,M_{\odot}}\right]^{\alpha},
\end{equation}

All is implemented in our SZ cluster counts module in the framework of the parameter inference Monte Python code \citep{Brinckmann:2018cvx}.
When running MCMC chains, we let mainly the normalisation parameter $(1-b)$ for SZ vary along with the six CMB cosmological parameters, the matter density $\Omega_m$, the~baryonic density $\Omega_b$, the~spectral index $n_s$, the~amplitude of the power spectrum $A_s$, the~Hubble constant $H_0$, and the optical depth reionization parameter $\tau_{reio}$, while also leaving $\alpha$ free since it could be degenerate with the calibration factor.

Since a modified value of the large-scale structures growth by mean of the growth index $\gamma$ we define next, enters the HMF through Equation~(\ref{eq:sigR}) and changes the cluster number counts, and~since this happens as well for a given value of $\sigma_8$ (defined as the $z=0$ variance in the density field at scales of $8h^{-1} \, {\rm Mpc}$ and used as a calibration parameter in (\ref{eq:sigR}), we therefore want to investigate in this work how the constraints on the cosmological parameters change when we allow both $\gamma$ and $\sigma_8$ to vary, where $\gamma$ affects the power spectrum as 
 \vspace{-6pt}
\begin{equation}\label{eq:pkmodgam}
    P_{\rm m}(k,z,\gamma) = P_{\rm m}(k,z) \left(\frac{D(z_*)}{D(z)} \frac{D(z,\gamma)}{D_(z_*,\gamma)} \right)^2
\end{equation}
where $\gamma$, the~growth index, is the parameter in the phenomenological parameterisation of the growth rate $ f=\Omega_{m}^\gamma(z)$ and $ f=d\ln \rm D / d\ln  a$,  $D$ being the growth of perturbations $\delta(z) = \delta_0  D(z)$ (see \cite{2023xnw} for another implementation of the effect of $\gamma$). The growth rate in $\Lambda$CDM is well approximated when the growth index is set to $\sim 0.545$ and takes different values in other modified gravity models~\cite{Linder:2007hg}.

While $\sigma_{8,f}$ used as free calibration parameter would change the power \linebreak spectrum~following
\begin{equation}\label{eq:pkmodsig8}
    P_{\rm m}(k,z,\sigma_{8,f}) = P_{\rm m}(k,z) \left(\frac{\sigma_{8,f}}{\sigma_{8,d}} \right)
\end{equation}
where $f$ stands for the value used for calibration while $d$ is for the derived one
\begin{equation}
    \sigma_{8,d} = \frac{1}{2 \pi^2} \int{{\rm d}k\,P_{\rm m}(k, z = 0) \left|W_{\rm TH}(k R_8)\right|^2 k^2},
  \label{equ:sigma8d}
\end{equation}
with $W_{\rm TH}(x) = 3(\sin{x} - x \cos{x})/x^3$ a top-hat filter in Fourier~space.

To better constrain the increase in the number of degree of freedom from our model independent approach we additionally combine datasets of the growth measurements $f(z)\,\sigma_8(z)$ obtained from the anisotropic clustering of galaxies after marginalising over the galaxy bias $b$. However, we need to correct for the Alcock--Paczynski (AP) effect or the altering of the distortions in the radial direction by the change of the Hubble parameter $H(z)$ with cosmology and in the transverse direction by that of the angular diameter distance $d_A(z)$.
Specifically, we implement the correction as follows~\cite{Arjona:2020yum}; first, we define the ratio of the product of the Hubble parameter $H(z)$ and the angular diameter distance $d_A(z)$ for the model at hand to that of the fiducial cosmology,
\begin{equation}\label{equ:APratio}
\textrm{ratio}(z)= \frac{H(z)d_A(z)}{H^{fid}(z)d_A^{fid}(z)}.
\end{equation}

 We then use it to correct the vector $V^i(z_i,p^j)$ that enters our likelihood, where $z_i$ is the redshift of $i$th point and $p^j$ is the $j$th component of a vector containing the cosmological parameters that we want to determine from the data, following
\begin{equation}
V^i(z_i,p^j)=f\sigma_{8,i}-\textrm{ratio}(z_i) f\sigma_{8}(z_i,p^j)
\end{equation}
where $f\sigma_{8,i}$ is the value of the $i$th datapoint, with~$i=1, \dots, N$, where $N$ is the total number of points, while $f\sigma_{8}(z_i,p^j)$ is the theoretical prediction, both at redshift $z_i$.\\

However, as~mentioned previously, this method performs a global correction, regardless of the shape of the power spectrum, its calibration and the galaxy matter bias. That is why~\cite{2016MNRAS.456.3743A} already considered a more sophisticated correction where the $f\sigma_8$ measurements in the `new' cosmology could be obtained from the fiducial one following
\begin{equation}
{f\sigma_8}_{new} = {f\sigma_8}_{fid} \, C \, \left(\frac{\alpha_\parallel}{\alpha_\perp^2} \right)_{new}^{(3/2)} \left(\frac{\sigma_8^{new}}{\sigma_8^{fid}} \right)^2,
\label{eqn:fs8}
\end{equation}
which is valid when $C$ below is considered $\sim 1$: 
\begin{equation}
C = \int_{k_1}^{k_2} dk \sqrt{\frac{P_{fid}^{m}(k)}{P_{new}^{m}(k')}}. 
\end{equation}

This correction was obtained starting from
\begin{equation}
\beta_{new} = \beta_{fid} \, C \, \frac{\mu_{fid}^2}{\mu_{new}^2} \sqrt{\frac{1}{\alpha_\parallel \alpha_\perp^2}} 
\label{eqn:beta2}
\end{equation}
where $\beta = f/b$ with $f$ the growth rate and $b$ the bias and $\mu$ the cosine of the angle of sight, and~where we needed to consider that  $\alpha_\parallel^2 \approx \alpha_\perp^2$ to obtain,
\begin{equation}
\frac{\mu_{fid}^2}{\mu_{new}^2} \approx \left(\frac{\alpha_\parallel}{\alpha_\perp}\right)^2,
\label{eqn:muratio}
\end{equation}
and be able to simplify $\mu$ from the equation. Finally, it needs also to consider that the bias measured is proportional to the $\sigma_8$ value. 

Here we keep the assumption on the bias but relax the other assumptions and try to correct the growth measurements by adjusting their values so that the integrated observed power spectrum in the fiducial cosmology remains equal to the one in the new cosmology varied with each step of the Monte Carlo chain method following the idea that the integrated observed power spectrum is a background evolution independent quantity. This method allows us also to incorporate the bias since it is a part of the observed power spectrum while in other studies the $f\sigma_8$ values are provided marginalising over it. For that we reconstituted the set of measurements choosing from each survey the couple of $f$ and $b$ (all compiled in Table~\ref{tab:fs8}), noting that it could be different from the $f$ value usually obtained when marginalising over the bias. We are by then trying to benefit from the effects of the variation of the cosmological parameters on the integrated full shape of the power spectrum, hoping that this constrain will serve to further reduce the degeneracy coming from the extra degrees of freedom considered in our model independent approach.

In practice, starting from the `fiducial' measurements we consider that the observed power spectrum, integrated over all angles and scales, should stay the same while the model parameters change,
\begin{equation}
C_{fid} = \int_{k_1}^{k_2} \int_{\mu_1}^{\mu_2} \left(b(z)\sigma_8(z)+f(z)\sigma_8(z)\,\mu^2\right)^2 \frac{P_{m}(k,z)}{\sigma_8^2(z)} \, d\mu \, dk,  
\end{equation}
 we equate with the new integrated observed power spectrum which components are denoted below by the prime symbol.
\begin{equation}
\int_{k'_1}^{k'_2} \int_{\mu'_1}^{\mu'_2} {\left(b'(z)\sigma'_8(z)+f'(z)\sigma'_8(z)\,\mu'^2\right)}^2 \frac{P'_{m}(k',z)}{\sigma'^2_8(z)} \, d\mu \, dk  = C_{fid}.
\end{equation}

We then solve for $f'\sigma'_8$ as function of $C_{fid}$ which is a known value, and $P'_m/\sigma'_8$ which is known in the new cosmology regardless of the value of $\gamma$, and $b'\sigma'_8$ also known assuming the bias is proportional to the modified $\sigma_8$ in the new set of~parameters.

\begin{table}[H]
	\caption{Compilation of the $f\sigma_8(z)$ and $b\sigma_8(z)$ measurements used in this analysis and \linebreak  related references.}
 \label{tab:fs8}
 	\newcolumntype{C}{>{\centering\arraybackslash}X}
		\begin{tabularx}{\textwidth}{CCCCC}
			\toprule
$\boldsymbol{z}$     & $\boldsymbol{f\sigma_8(z)}$ & $\boldsymbol{\sigma_{f\sigma_8(z)}}$&$\boldsymbol{b\sigma_8(z)}$  & \textbf{Ref.} \\
\midrule
0.15    &     0.49 & 0.145 & 1.445 & \cite{Howlett:2014opa}\\
0.18	&	0.392	&	0.096	&	0.894	&	\cite{Blake:2013nif}		\\
0.38	&	0.528	&	0.072	&	1.105	&	\cite{Blake:2013nif}	\\	
0.25	&	0.3512	&	0.0583	&	1.415	&	\cite{Samushia:2011cs}	\\	
0.37	&	0.4602	&	0.0378	&	1.509 &	\cite{Samushia:2011cs}	\\	
0.32	&	0.394	&	0.062	&	1.281	&		\cite{Gil-Marin:2015sqa}\\	
0.57	&	0.444	&	0.038	&	1.222	&		\cite{Gil-Marin:2015sqa}\\
0.22	&	0.42	&	0.07	&	0.664	&	\cite{Blake:2011rj}\\	
0.41	&	0.45	&	0.04	&	0.728	&	\cite{Blake:2011rj}	\\	
0.6	&	0.43	&	0.04	&	0.880	&	\cite{Blake:2011rj}\\
0.78	&	0.38	&	0.04	&	0.973	&	\cite{Blake:2011rj}\\	
0.6	&	0.55	&	0.12	&	0.730	&	\cite{Pezzotta:2016gbo}	\\	
0.86	&	0.4	&	0.11	&	0.740	&	\cite{Pezzotta:2016gbo}	\\
1.4	&	0.482	&	0.116	&	0.814	&	\cite{Okumura:2015lvp}	\\
0.978	&	0.379	&	0.176	&	0.826 &	\cite{Zhao:2018gvb}	\\	
1.23	&	0.385	&	0.099	&	0.894 	&	\cite{Zhao:2018gvb}	\\	
1.526	&	0.342	&	0.07	&	0.953 	&	\cite{Zhao:2018gvb}	\\	
1.944	&	0.364	&	0.106	&	1.080 	&	\cite{Zhao:2018gvb}	\\	
			\bottomrule
		\end{tabularx}
\end{table}

\section{Results}\label{sect:results}

\begin{figure}[H]
	\hspace{-3pt}\includegraphics[width=12.5cm]{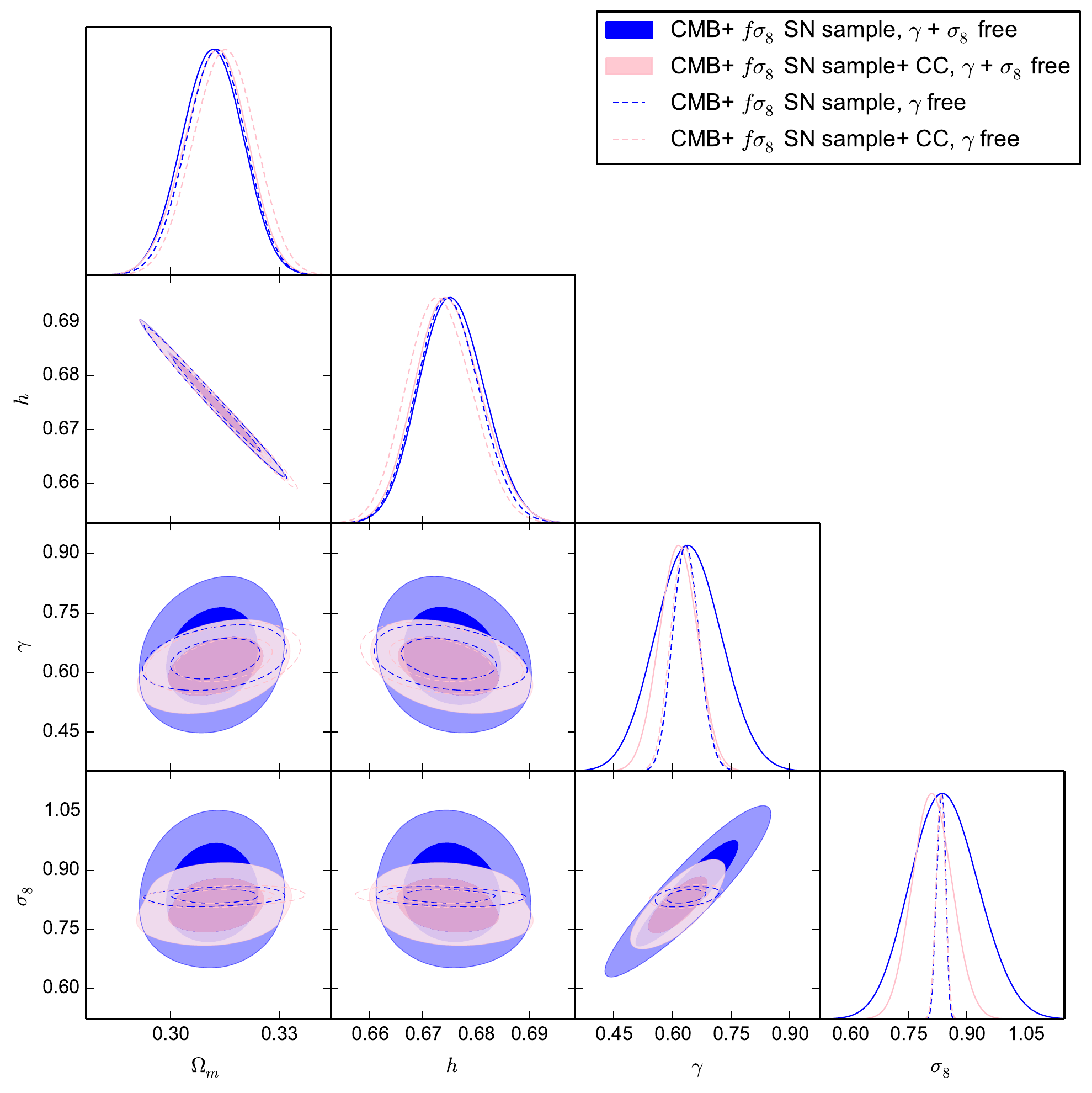}
    \caption{The 68\% and 95\% confidence contours for the parameters $\Omega_m$, $h$, $\gamma$ and a derived $\sigma_8$, all inferred from a combination of CMB $C_{\ell}^{TT,TE,EE}$ Planck 2018, $f\sigma_8$ measurements from \cite{2018PhRvD..98h3543S} and SZ detected cluster counts (dashed lines) in comparison to those with $\sigma_8$ as a free parameter using the same~probes.}
    \label{fig:s8vss8free}
\end{figure}

We start by showing in Figure~\ref{fig:s8vss8free} the impact on the MCMC inferred constraints from considering both $\gamma$ and $\sigma_8$ as a free parameter in comparison to those obtained when the growth index is free but $\sigma_8$ is derived using Equation~(\ref{equ:sigma8d}). When adopting the first prescription, $\sigma_{8,f}$ becomes effectively the one driving the observable theoretical prediction. That is why we show both, the derived and free one, on the same plot.  We also show the results from combining the CMB and growth measurements data in comparison to the case when further adding cluster counts constraints. Here we use growth measurements compiled by~\cite{2018PhRvD..98h3543S} and follow the global correction of Equation~(\ref{equ:APratio}) to account for the AP effects. As expected, the constraints are tighter when less free parameters are considered. We also observe that when $\gamma$ is the only free parameter, the constraints from CMB + $f\sigma_8$ do not show substantial changes with respect to when we add cluster counts because the growth is already fixed by the CMB and $f\sigma_8$ combination while leaving $\gamma$ and $\sigma_8$ free to add a degeneracy that needs the contribution of the cluster constraints to be broken, which results in tighter contours with respect to the case of using CMB + $f\sigma_8$ only. While the derived $\sigma_8$ value is compatible with that usually inferred from CMB, considering $\sigma_8$ as a free parameter widens the constraints, allowing a range that covers values for $\sigma_8$ that are usually obtained from either deep or local probes. It permits by then to relieve the $\sigma_8$ discomfort essentially because it also allows larger bounds on $\gamma$ though still staying compatible with 0.55 the $\Lambda$CDM value; while in the case where only $\gamma$ is left free, the~latter is restricted from exploring values that reconcile CMB and cluster counts measurements, and that remains true even if we additionally combine with cluster counts. This happens albeit from a tightening in the constraints in the latter case due to the fact that even if $\sigma_8$ is not fixed by CMB anymore, however the latter is needed to constrain the matter density while the combination of $f\sigma_8$ + CC will constrain and break the degeneracy on the growth and the normalisation of the matter fluctuation of parameter on local redshifts all~together.


\begin{figure}[H]
	\includegraphics[width=\textwidth]{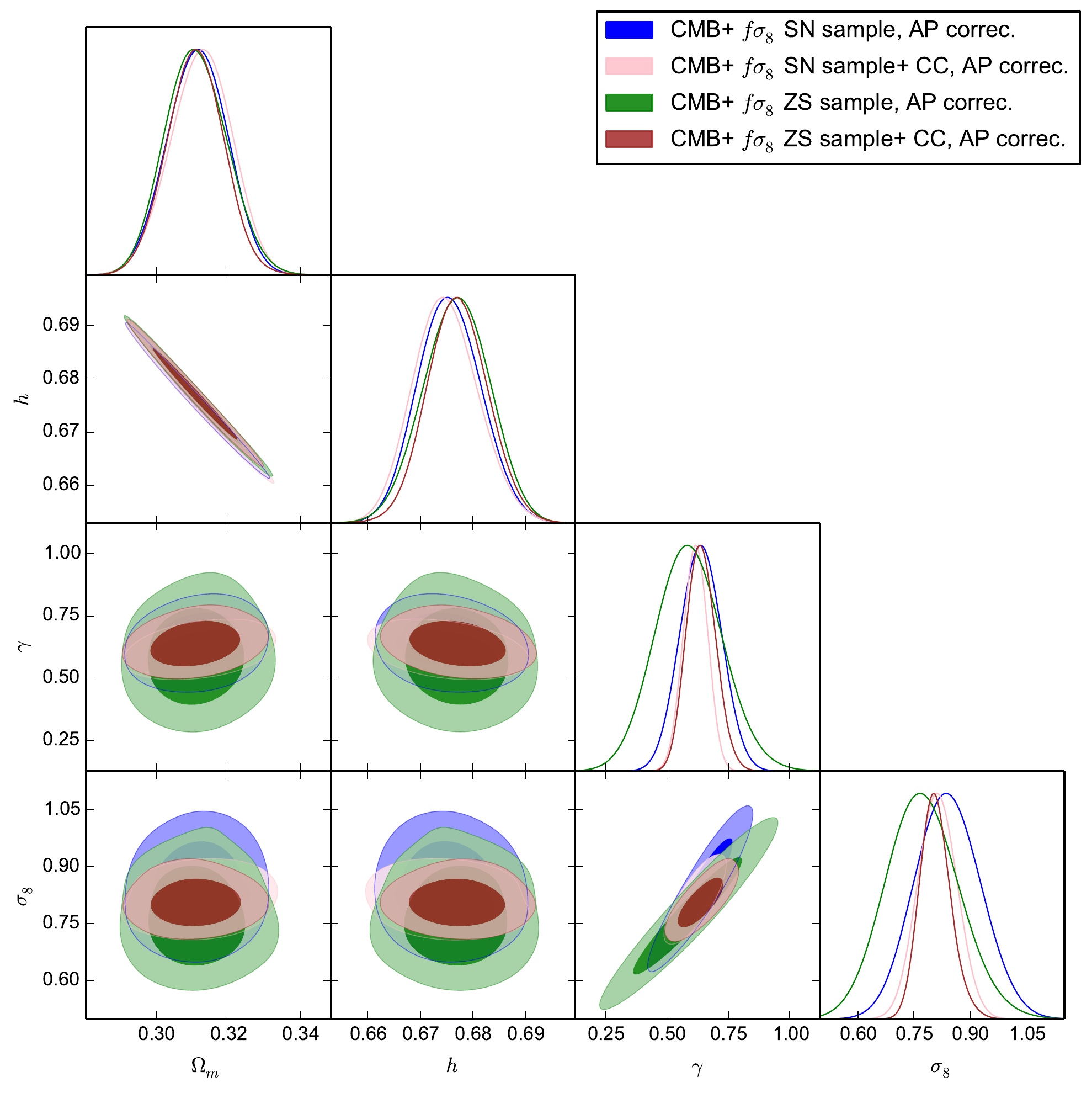}
    \caption{The 68\% and 95\% confidence contours for $\Omega_m$, $h$, $\gamma$ and a free $\sigma_8$, all inferred from a combination of CMB $C_{\ell}^{TT,TE,EE}$ Planck  2018, SZ detected cluster counts and $f\sigma_8$ measurements from \cite{2018PhRvD..98h3543S} corrected by rescaling the final observation for the AP effect, in~comparison to those inferred using the same probes but using instead $f\sigma_8$ measurements from Table~\ref{tab:fs8} using the same correction~method.}
    \label{fig:s8vss8AP}
\end{figure}

Before we discuss the impact of the correction that relies on adjusting the growth to preserve the integrated observed power spectrum for the AP effect rather than only performing a global correction, we want to check whether the new compilation needed to perform the new correction is compatible with the standard previous compilation used. For that we show in Figure~\ref{fig:s8vss8AP} a comparison between the constraints inferred from the two compilations using the same classical AP correction. The~two constraints are compatible despite the difference in some of the datasets that were omitted or added with respect to the old compilation and despite the fact that some values for $f\sigma_8$ differs even for datasets that are in common between both compilations, since in some of them we considered the $f\sigma_8$ obtained without marginalising over the bias. Here, we only show the sufficient extreme case of letting both $\gamma$ and $\sigma_8$ free. We also notice that the agreement is bigger when we add subsequently cluster counts since, as expected, the latter further limit the shift or the widening of the contours induced from choosing different values for $f\sigma_8$ in the two compilations.

\begin{figure}[H]
	\hspace{-3pt}\includegraphics[width=\textwidth]{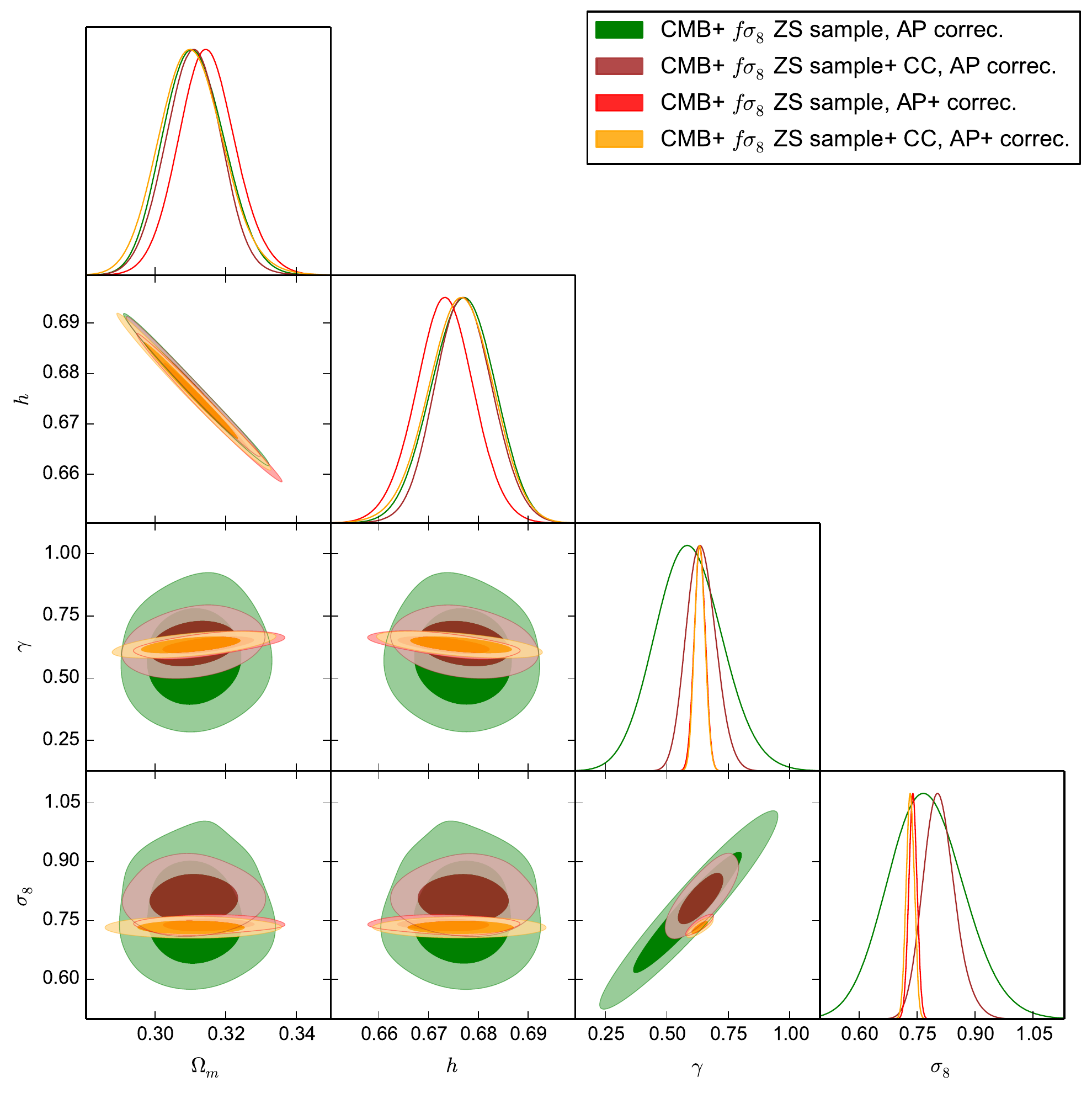}
    \caption{The 68\% and 95\% confidence contours for $\Omega_m$, $h$, $\gamma$ and a free $\sigma_8$, all inferred from a combination of CMB $C_{\ell}^{TT,TE,EE}$ Planck 2018, SZ detected cluster counts and $f\sigma_8$ measurements from \cite{2018PhRvD..98h3543S} corrected by rescaling the final observation for the AP effect, in~comparison to those inferred using the same probes but using instead $f\sigma_8$ measurements from Table~\ref{tab:fs8} corrected by preserving the integrated power spectrum used to obtain the growth measurements for the AP effect change.}
    \label{fig:s8SNvss8ZS}
\end{figure}

Having established the compatibility between the two combination of datasets we now show in Figure~\ref{fig:s8SNvss8ZS} the impact in the $\gamma$ + $\sigma_8$ free case when considering the new AP correction on the integrated power spectrum in comparison to the global one. We observe that the correction introduced makes the model more stiff and by then strongly tighten the contours, notably on our two parameters $\gamma$ and $\sigma_8$. The~former is then constrained tightly around $\sim$0.66 without forbidding the latter from showing preferences for values compatible with cluster counts. We also notice that the improvement from adding cluster counts is small here since the correction boosts the constraints from $f\sigma_8$ while cluster counts are not affected by~it.

\begin{figure}[H]

\centering
\includegraphics[width=\textwidth]{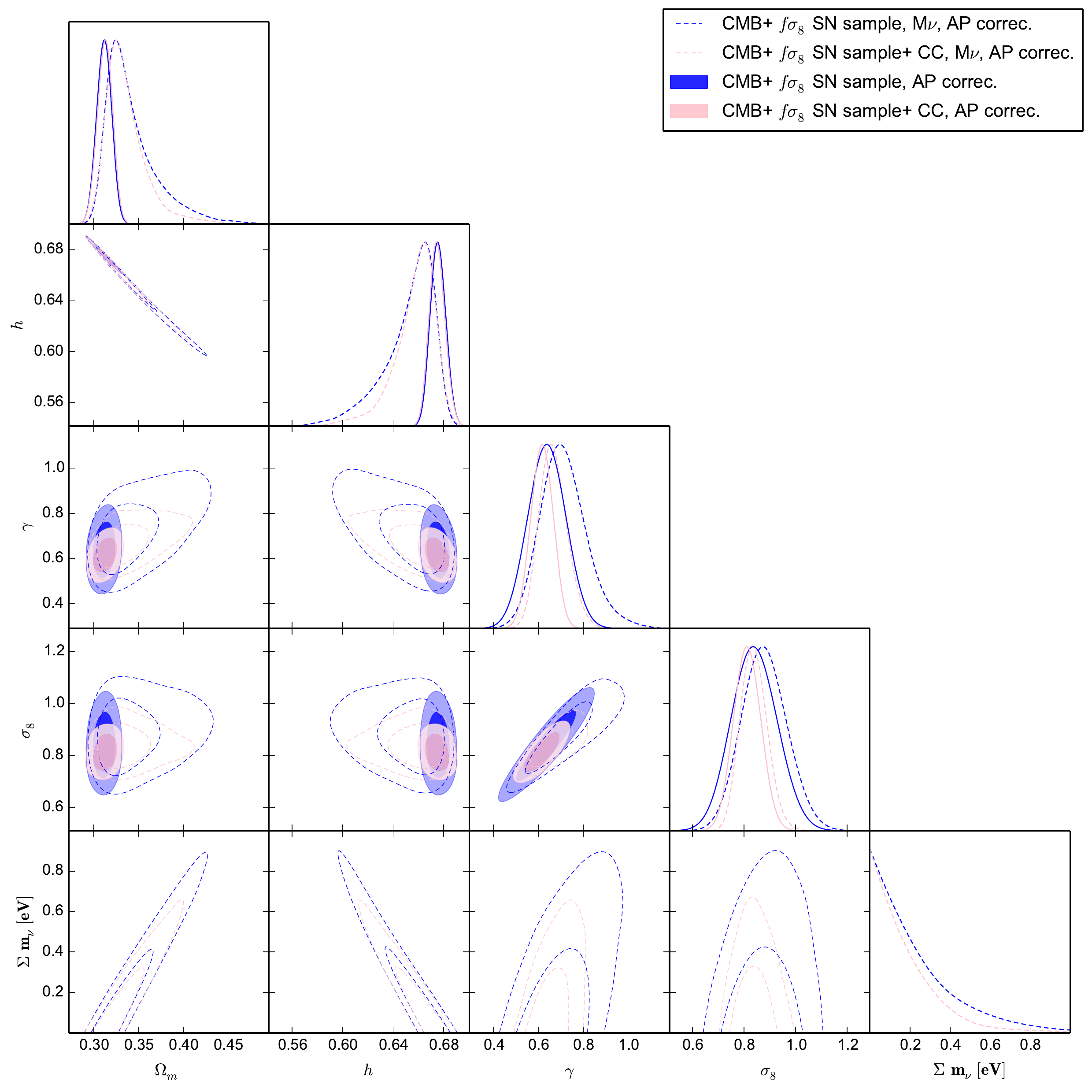}
    \caption{The 68\% and 95\% confidence contours for $\Omega_m$, $h$, $\gamma$ and a free $\sigma_8$, all inferred from a combination of CMB $C_{\ell}^{TT,TE,EE}$ Planck  2018, SZ detected cluster counts and $f\sigma_8$ measurements from \cite{2018PhRvD..98h3543S} corrected by adjusting the growth of the final observation for the AP effect for the case of massless neutrino in comparison to when considering massive~neutrinos.}
    \label{fig:s8vss8nuSN}
\end{figure}

\begin{figure}[H]
	\includegraphics[width=\textwidth]{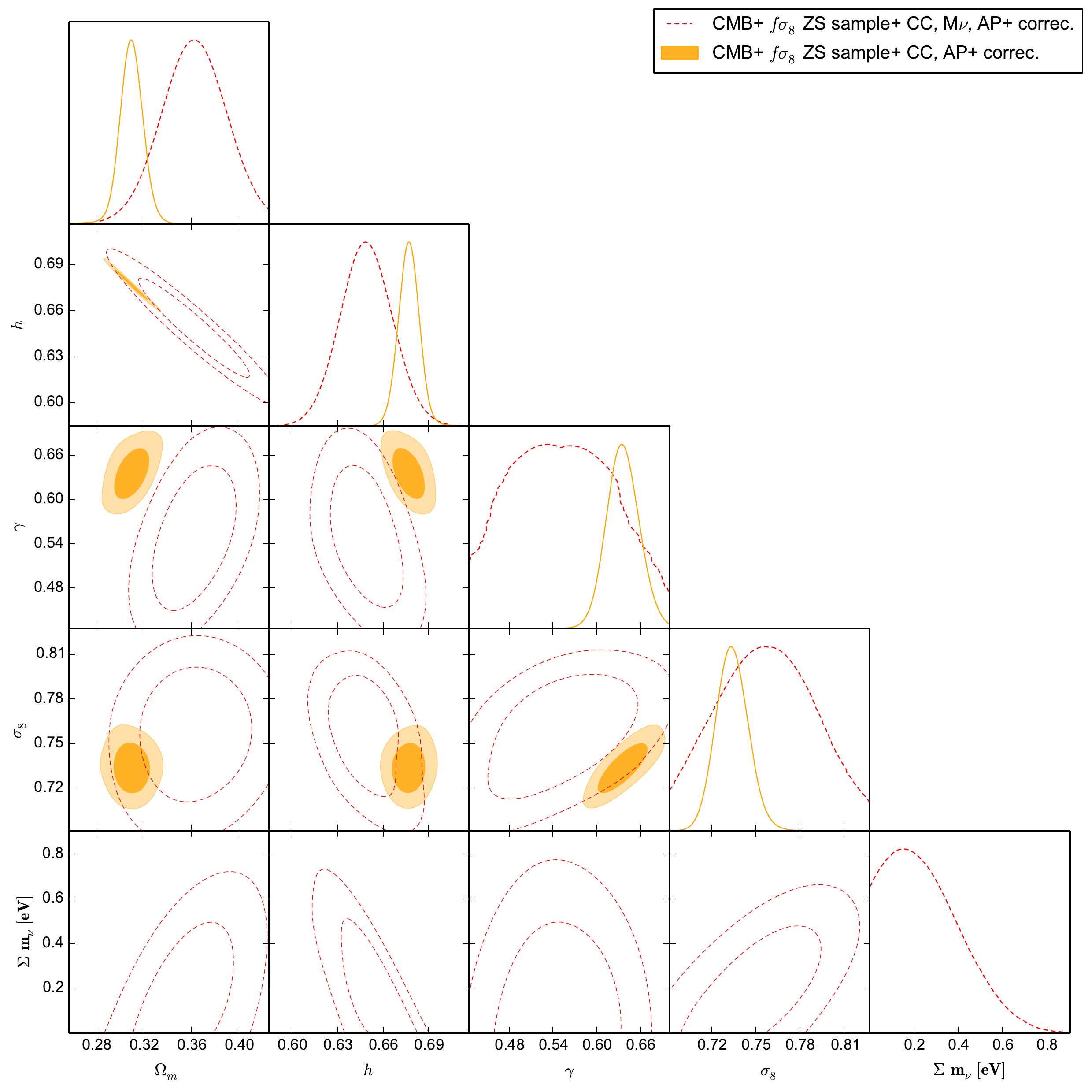}
    \caption{The 68\% and 95\% confidence contours for $\Omega_m$, $h$, $\gamma$ and a free $\sigma_8$ with massive neutrinos, all inferred from a combination of CMB $C_{\ell}^{TT,TE,EE}$ Planck 2018, SZ detected cluster counts and $f\sigma_8$ measurements from \cite{2018PhRvD..98h3543S} corrected by rescaling the final observation for the AP effect in comparison to $f\sigma_8$ measurements from Table~\ref{tab:fs8} corrected by preserving the integrated power spectrum, used to obtain the growth measurements, from the AP effect change.}
    \label{fig:s8vss8nuZS}
\end{figure}

Finally, massive neutrinos have been proposed to solve the $\sigma_8$ tension due to the fact that they free stream from the halo gravitational potentials lowering by then the power spectrum. However, \cite{Sakr:2018new} has shown that they are not able, especially when CMB+cluster counts are further combined with BAO, to fix the tension on $\sigma_8$ even if we further allow $\gamma$ to vary. Here we also consider a case with free massive neutrinos in our two schemes with $f\sigma_8$ data but without adding BAO constraints. However, we also allow, as above, $\sigma_8$ and $\gamma$ to vary with massive neutrinos using first the usual AP correction, in which the neutrinos impact enters from its effects on the background evolution and translates into those on $d_A$ and $H(z)$ in Equation~(\ref{equ:APratio}). We observe in Figure~\ref{fig:s8vss8nuSN} that introducing neutrinos have an impact this time on $h$ and $\Omega_m$
 parameters skewing both to large and smaller values, respectively, with however each effect compensating the other. This is also seen from the observed correlation between $h$ and $\Omega_m$, resulting in a small effect on $\gamma$ and even smaller on $\sigma_8$, suggesting that even with more degrees of freedom, neutrinos do not have a substantial impact on solving the tension. However, if we consider now our more tiered correction with the same free parameters, we observe in Figure~\ref{fig:s8vss8nuZS} that allowing massive neutrinos relaxes the stiffness of the correction we previously considered and allow for larger bounds on all parameters while leaving the constraints on $\sigma_8$ compatible with cluster counts, the same as it was the case with massless neutrinos. We also notice that $\gamma$ is kept within its $\Lambda$CDM values at the expense of showing a small preference for a non-vanishing value for its mass. This also should call for attention to be taken when usually assessing the bounds on neutrinos from growth or structure formation observations using the classical AP correction.\\\\\\\\

\vspace{-6pt}

\section{Conclusions}\label{sect:conclusion}

In this work, we performed a Bayesian study to obtain constraints on the cosmological parameters, in~particular the matter fluctuation parameter $\sigma_8$ when additionally varying the growth index $\gamma$, and~that using three probes, the~CMB angular power spectrum, RSD from galaxy clustering and galaxy cluster abundances. Usually, due to the interdependency between the effects of our two parameters, two of the aforementioned three probes are sufficient to constrain them. However, here we further tried a model independent approach, by introducing two additional degrees of freedom by means of a parameter that rescales the value of $\sigma_8$, as well as by letting free the mass-observable calibration parameter for the cluster counts probe. Hence the combination of the three probes is needed to break the degeneracy and reduce the degrees of freedom from our model-independent approach. Moreover, this is further motivated by the existence of a small discrepancy between the value of $\sigma_8$ obtained from cluster counts with respect to that inferred from CMB data that could as well be degenerate with the growth of structure and by then $\gamma$. Since it was already found in~\cite{Ilic:2019pwq} that the combination of CMB and SZ detected cluster counts in the presence of a free growth index $\gamma$ is barely able of alleviating the discrepancy even if we also relax the mass-observable calibration parameter, due to the tomographic constraints put on the growth at different redshifts, we are expecting that the further combination with $f\sigma_8$ data would further constrain $\gamma$ from fixing the discrepancy; however, this was obtained with a model dependency relating the calibration of the amplitude of the CMB power spectrum to the nowadays value of $\sigma_8$ through the growth rate parameterised by $\gamma$ while here, by further considering $\sigma_8$ as a free and not as a derived parameter, we expect to further relax the previous constraints in the hope of reducing or fixing the tension. 

We found that allowing a free $\sigma_8$ does indeed widen the constraints on the latter if the $f\sigma_8$ measurements are combined with CMB and to a lesser but still substantial level if we further add cluster counts data, and~we found that the growth index bounds are also still in agreement with the equivalent $\Lambda$CDM values. However, the latter improvement with respect to fixing the tension was obtained when using the common global AP geometrical correction usually performed to the growth measurements to account for every new set of parameters explored by the MCMC inference method. When we tried a more sophisticated correction based on preserving the integrated observed power spectrum used in the growth measurements studies, we found that the constraints were tightened again but the discrepancy was fixed through a shift towards the local values for $\sigma_8$ while the growth index is in slight disagreement with its $\Lambda$CDM value. Allowing further free massive neutrinos, an ingredient often advocated as a solution to fix the discrepancy; though this belief was disfavoured by \cite{Sakr:2018new}; slightly reduces the discrepancy, and that by relaxing the constraints in all cases including the most tightening ones when the three probes CMB, $f\sigma_8$ and CC are combined, all when adopting the classical AP global correction. However, when using the more tiered correction, we observe that allowing massive neutrinos relatively relaxes more the constraints with respect to the case with massless neutrinos, with values of $\sigma_8$ still in agreement with its local ones while $\gamma$ is now within its $\Lambda$CDM fiducial values. However, in the latter case, neutrinos masses are now showing a small preference for non vanishing values.

We conclude that untying the growth index (a parameter encapsulating deviations from $\Lambda$CDM) relation with $\sigma_8$ helps in reducing the tension, and that careful analysis of the data products should be taken to limit biases from the model-dependency in their treatment, in order to better accurately asses the discrepancy's significance and its impact on the viability of the $\Lambda$CDM model.

\vspace{6pt}

\dataavailability{All datasets used in this research are public and available from the cited references.} 

\acknowledgments{Z.S. acknowledges funding from DFG project 456622116 and support from the IRAP Toulouse and IN2P3 Lyon computing centres.}

\conflictsofinterest{The authors declare no conflicts of interest.}

\begin{adjustwidth}{-\extralength}{0cm}

\reftitle{References}



\bibliographystyle{mdpi}

\PublishersNote{}
\end{adjustwidth}
\end{document}